# Impartial binary decisions through qubits

**Sujan Vijayaraj [1], S. Nandakumar [1]**

**Abstract**

Binary decisions are the simplest form of decisions that are made in our daily lives. Examples include choosing a two-way path in a maze, accepting or declining an offer, etc. These decisions are also made by computers, machines and various electronic components. But decisions made on these devices can be partial and deterministic, and hence compromised. In this paper, a simple framework to implement binary decisions using one or many qubits is presented. Such systems are based on a separate hardware infrastructure rather than computer codes. This helps enable true randomness and impartial decision making. The multi-armed bandit problem is used to highlight the decision making ability of qubits by predictive modelling based on quantum Bayesianism. Bipartite and multipartite entangled states are also used to solve specific cases of the problem.

## 1   Introduction

Decision making in uncertain environments with regards to various applications can be subject to bias and ad hoc assumptions in the absence of an impartial decision maker (Beach et al. 2017). Applications may include equitable allocation of communication services to users, generating random binary numbers for cryptographic applications, financial decisions, etc. The underlying probability factor or the extent of randomness is not truly captured due to determinism in classical devices. For instance, pseudo random number generation codes which are expected to be secure, can be generated after a certain period by a third party if one of the bits or the initial bit in the random sequence is known. This process is known as seeding. Though several stochastic processes are used to generate random numbers in between the sequence to avoid seeding, the associated classical devices are still not impervious to security threats. For example, the Mersenne Twister algorithm which produces a pseudo random sequence uses a computer's real time clock to avoid seeding. For this reason, the search of true random devices like quantum random number generators and impartial decision makers have generated a lot of interest.

   A qubit or a quantum bit is the analogue to the classical bit in quantum information processing. A classical bit can have a value of 0 or 1. But a qubit can be expressed as a linear superposition of two fundamental states. These states could be the horizontal/vertical polarization of a single photon, top/down spin of an electron, etc. We use the horizontal/vertical polarization of single photons in this paper. Other polarization basis also exist but we will confine ourselves to the horizontal (0°) and vertical (90°) basis. Following this idea, a photon can be expressed as a linear superposition of horizontal and vertical states $\alpha|0> + \beta|1>$ where $\alpha, \beta \in \mathbb{C}$ and $|\alpha|^2 + |\beta|^2 = 1$, and $|0>$ and $|1>$ represent the horizontal and vertical states respectively (Nielsen and Chuang 2000). Unfortunately these qubits are fragile and would also collapse to a classical bit on measurement. Here $|\alpha|^2$ and $|\beta|^2$ are the probability amplitudes and are equal to the probability of resulting in the $|0>$ and $|1>$ state respectively after measurement.

---

Sujan Vijayaraj
sujankvr@gmail.com

S. Nandakumar
snandakumar@vit.ac.in

1 School of Electronics Engineering,
   Vellore Institute of Technology, Vellore, India

We explain how qubits are random and present a framework as to how they can be used to make impartial, non-deterministic binary decisions through the multi-arm bandit problem. The discussion starts with the single agent problem and is later extrapolated to multiple users. The multi-arm bandit problem helps maximize rewards in social settings involving one or many players by making appropriate decisions. Such a problem is being successfully applied to a wide array of fields (Mahajan et al. 2008). Fundamental quantum properties like superposition and entanglement are used to solve the problem. The probability amplitudes of the qubits involved are updated sequentially to establish a predictive model based on quantum Bayesianism. The approach seeks to implement hardware exclusively for decision making unlike conventional algorithms.

## 2  Intrinsic randomness of qubits

From the expression of a qubit, one can notice how probability amplitudes can affect the outcome of the measurement. This can open avenues for true randomness. For instance, let $|\alpha|^2 = |\beta|^2 = 0.5$ in the expression of the qubit state. The corresponding state is $(|0> + |1>)/\sqrt{2}$, which can also be denoted as $|+>$. Such a state can be produced by passing a single photon through a polarizer aligned at 45°. Further a balanced polarization beam splitter (PBS) can be used to split the horizontal and vertical components, and at each of these outputs, a detector is placed. Since a single photon is used, only one of the detectors can detect a photon at a time. The detection of horizontally and vertically polarized photons can be regarded as the 0 and 1 bit respectively. The detectors follow the randomness of the generated qubit and hence true randomness is achieved as opposed to deterministic bits generated by pseudo random number generators (Jennewein et al. 2000). Such devices are called quantum random number generators. Another quantum random number generator uses transition of photons from a single mode fiber to two different multi-mode fibers (Stefanov et. al 2000). There is a 50:50 probability for the photons exiting the single mode fiber to travel through each of the multi-mode fibers. A delay is introduced in one of the multi-mode fibers to distinguish between the paths, and the two paths are labeled as 0 and 1 bit. Quantum random number generators based on vacuum fluctuations, laser phase noise, Raman scattering, etc. have been proposed over the years catering to high speed applications (Herrero-Collantes et. al). In these examples, the probability amplitudes were equal to achieve true randomness. What would happen if the probability amplitudes are unequal?

## 3  Binary decisions and the single agent multi-armed bandit problem:

Presently, important complex decision making systems suffer from algorithmic bias (Hajian et al. 2016). Apart from using further algorithmic techniques to tackle the problem, it is important to explore decision making devices that are detached from classical computation techniques which are based on Bayesian probability theory. This makes quantum decision making a worthwhile alternative (Patra 2019).

In a binary decision, each event can take up the notation of 0 and 1. Hence if all factors are accounted for assigning the probability amplitudes of a qubit, the qubit measurement can yield an impartial non-deterministic result. If a user requires 0 to occur in a sequence 60 % of the time, the qubit state $\sqrt{0.6} |0> + \sqrt{0.4} |1>$ can be generated by appropriately rotating the polarizer to accommodate the skew. Repeated generation and measurement of the state can generate the intended sequence. Hence binary decisions can remain impartial.

The single agent multi-armed bandit problem is based on making such binary decisions. For a simple case, consider two slot machines from which a user can use one machine at a time. The slot machines have an inbuilt probability distribution for reward which the user doesn't know. Hence the user must use the machines to arrive at a better understanding of the rewards. More the usage, better the rewards. But excessive use would also lead to increased losses. The user must navigate this exploration-exploitation dilemma to maximize rewards. The problem has several real world applications from management to communication technology to healthcare. Naruse et al. (2005) experimentally investigated the problem using single photons from nitrogen vacancies in a nano diamond as qubits.

Since this is a reinforcement learning problem, bias should favor the machine which rewards the user. Let the two machines be M0 and M1, and let the 0 and 1 bit denote the event of M0 and M1 rewarding the user respectively.

The measurement apparatus with the PBS which is explained above is considered for simplicity. Now the users can start from the assumption that the probabilities of winning from each machine is equal. Hence the user can generate a $|+>$ state initially and measure the state. If the result is 0, the user will use M0; else if it is 1, the user uses M1. If M0 rewards the user, the polarizer can be rotated towards the horizontal such that $|\alpha|^2$ becomes $|\alpha|^2 + c$, where c is a real constant and greater than zero (all constants in the paper follow the same condition unless otherwise stated). Incidentally $|\beta|^2$ becomes $|\beta|^2 - c$. If M0 doesn't reward the user, the decision is taken in favor of M1 by rotating the polarizer towards the vertical such that $|\alpha|^2$ becomes $|\alpha|^2 - c$ and $|\beta|^2$ becomes $|\beta|^2 + c$. The situation is reverse in case of M1. The value of the constants should be low enough so that the coefficients do not get close to unity very soon. At the same time, very low values would not converge easily. In general, the new qubits are $\sqrt{|\alpha|^2 + c}\,|0> + \sqrt{|\beta|^2 - c}\,|1>$ or $\sqrt{|\alpha|^2 - c}\,|0> + \sqrt{|\beta|^2 + c}\,|1>$. The next iteration takes place with these new qubits. If the probabilities of the systems rewarding the user P1 and P2 remain constant, the resulting qubit after many iterations would be asymptotically equal to

$$\frac{P1}{P1+P2}|0> + \frac{P2}{P1+P2}|1>.$$

Note that the qubits are totally non-deterministic and impartial predictive models, and would converge eventually to understand the probability distribution of reward of the system.

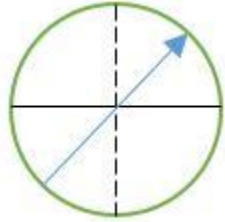

**Fig 1.** If the arrow is the polarizing angle, it is rotated towards the horizontal (normal line) when a 0 bit is measured. If a 1 bit is measured, it is rotated towards the vertical (dashed line).

## 4  Entangled states for competitive multi-armed bandit problems

Let's consider two users U and V in the same setup described in the previous section. Now this becomes a multi-armed bandit problem which can give rise to conflicts. The best way to remove conflicts and maximize rewards is by generating and measuring a maximally entangled state $(|0_U 1_V> + |1_U 0_V>)/\sqrt{2}$ each time before using the machines, as described experimentally by Chauvet et al. (2019). Orthogonal qubits have been used because the users must use distinct machines in each iteration to remove conflict. The subscripts U and V denote the corresponding bits after the users' measurement. The entangled state can be generated from SPDC where pump photons pass through a polarizer aligned at 45° before striking a Type-2 non-linear crystal. This state produces a random 50: 50 output which is fair for both users because the reward probability of the machines are unknown and fluctuating. The qubit measurement must be done in the same basis for entangled states. It is easier if only one basis is used for the whole process. If the polarizer is aligned between 0° and 90°, but other than 45°, non-maximally entangled states are produced (White et al. 1999). But this enables bias in the procedure. The polarizer must be at 45° to maintain impartiality and should be refereed by both the users or by an unbiased arbitrator.

Let's take the case of two pairs of machines at different locations, with the same probability distribution for reward. It is not difficult to perceive the particular case as appropriate for many important applications – navigating the noise of similar fiber lines from the same manufacturer for example. The probability amplitudes of the qubits follow the same framework as described in Section 3. Assume machines M0, M1 can be accessed by user U and N0, N1 by user V where M0 and N0 follow the same probability distributions and likewise for M1 and N1. U and V are supposed to find the better rewarding machine. This problem naturally requires collective effort without any hostility for maximizing rewards as the users must also be aware of the other pair of machines to better learn the reward probability. Hence user U or V can handle the polarizer to generate entangled states. After generating the state, one of the qubits is sent to the other user through a quantum channel. Then they both make a measurement in the same basis. 0 and 1 bit denote the event of M0, N0 and M1, N1 rewarding the users respectively. Initially a maximally entangled state, $(|0_U 0_V\rangle + |1_U 1_V\rangle)/\sqrt{2}$ is generated. The qubits are non-orthogonal because they should use the similar machines at the same time. Let the measurement result be $|00\rangle$. It means U and V use M0 and N0 respectively. If both receive awards, the polarizer should be rotated towards the horizontal similar to the single agent multi-bandit problem. In general the resulting state would be $\sqrt{|\alpha|^2 + c}\, |0_U 0_V\rangle + \sqrt{|\beta|^2 - c}\, |1_U 1_V\rangle$. If both do not receive awards, the state $\sqrt{|\alpha|^2 - c}\, |0_U 0_V\rangle + \sqrt{|\beta|^2 + c}\, |1_U 1_V\rangle$ would be generated. If only one of the users is rewarded, the state is not altered and used again for measurement. This repeats sequentially with the new probability amplitudes. These qubits can also be non-maximally entangled unlike the situation described in the first paragraph of this section.

Appropriate Greenberger–Horne–Zeilinger (GHZ) states (multipartite systems, i.e. more than two qubits) can be used to increase the number of users. For example, if there is an additional user W, $(|0_U 0_V 0_W\rangle + |1_U 1_V 1_W\rangle)/\sqrt{2}$ is the initial state used. This state can be generated using the experiment described by Bouwmeester et al. (1999). If all three users receive the same measurement, the resulting quantum states would be $\sqrt{|\alpha|^2 + c_1}\, |0_U 0_V 0_W\rangle + \sqrt{|\beta|^2 - c_1}\, |1_U 1_V 1_W\rangle$ or $\sqrt{|\alpha|^2 - c_1}\, |0_U 0_V 0_W\rangle + \sqrt{|\beta|^2 + c_1}\, |1_U 1_V 1_W\rangle$ where $c_1$ is a constant. If only two users receive the same measurement, the resulting quantum state would be $\sqrt{|\alpha|^2 + c_2}\, |0_U 0_V 0_W\rangle + \sqrt{|\beta|^2 - c_2}\, |1_U 1_V 1_W\rangle$ or $\sqrt{|\alpha|^2 - c_2}\, |0_U 0_V 0_W\rangle + \sqrt{|\beta|^2 + c_2}\, |1_U 1_V 1_W\rangle$ where $c_2$ is a constant and $c_2 < c_1$. In general, if $n$ users are present, the $n$ qubit quantum states involved would be $\sqrt{|\alpha|^2 \pm c_i}\, |0_U 0_V 0_W ....\rangle + \sqrt{|\beta|^2 \mp c_i}\, |1_U 1_V 1_W ....\rangle$ where $i = 1, 2, 3.......[n/2]$, $c_i$ is a constant, and [ ] indicates the greatest integer function. This is to accommodate the varying probability of rewards depending on the results of the measurement. If there are $n$ users, $[n/2]$ constants are involved. The decision is made only when majority of the users get the same result. The constants are prepared before the users engage with the machines and are applied sequentially according to the rewards. As the number of users increase, the qubits can model the system more effectively. In spite of this advantage, one must keep in mind that practical realization of multipartite entanglement is difficult.

## 5   Conclusion

The paper elucidates how qubits can be used for future devices which rely on simple decision making. The framework can be applied on various cases of decision making other than the ones mentioned in the paper with a similar probability updating procedure. The underlying motivation to rely on such devices is to achieve true randomness and impartial probability based decisions. The famous multi-arm bandit problem is used to explore the dilemma of exploration-exploitation and how decisions can be made accordingly by bipartite and multipartite systems using fundamental quantum properties like superposition and entanglement. The constants that are used with the probability amplitudes are yet to be quantified with accordance to various experimental techniques.